# Physarum machine: implementation of Kolmogorov-Uspensky machine in biological substrate[*]

Andrew Adamatzky

Faculty of Computing, Engineering and Mathematical Sciences,
University of the West of England,
Bristol BS16 1QY, United Kingdom
andrew.adamatzky@uwe.ac.uk

April 8, 2007

### Abstract

We implement Kolmogorov-Uspensky machine on a plasmodium of true slime mold *Physarum polycephalum*. We provide experimental findings on realization of the machine instructions, illustrate basic operations, and elements of programming.

*K*eywords: nature inspired computing, theory of computation, plasmodium, theory of algorithms

## Introduction

Unconventional computing [37, 4, 13] is an interdisciplinary field of science, where computer scientists, physicists, mathematicians, apply principles of information processing in natural systems to design novel computing devices and architectures.

What is a typical piece of research in unconventional, or non-classical computation? Someone picks up, very often arbitrarily, physical, chemical or living system $S$, and interprets dynamics of $S$ in terms of computation. Universality of $S$ is usually proved by showing that $S$ can simulate functionally complete logical gates or a Turing machine. The approach is feasible indeed if $S$ has an intrinsic regularity or may be even a cellular-automaton structure inbuilt [2] but

---

may cause troubles when used in formalizations of chemical, morphogenetic and other amorphous computers with non-fixed architecture.

There is however an alternative to Turing machine — Kolmogorov-Uspensky machine (KUM) [24, 25] that perfectly reflects phenomena of growth and structure-dependent computation in natural systems. As Gurevich and Blass argue, the rational behind Turing machine was to automatize computation as it performed by a human, while KUM depicts "computation as a physical process" [11]. In present paper we implement KUM in biological substrate, vegetative stage of true slime mold *Physarum polycephalum*.

# 1 Kolmogorov-Uspensky machines (KUM)

In early 1950s Kolmogorov [24], later with his at that time student Uspensky [25], outlined a concept of an abstract machine defined on a dynamically changed graph-based structure, called 'Kolmogorov complex'. The finite undirected connected graph has nodes labeled in such manner that any two closest neighbors of any node has different labels and there is only active node at any step of development (neighborhood of any active node is fixed for any particular algorithm) [41].

KUM is a mother of all present models of real-life computation. From its inception in 1950th, it was reincarnated in 1968 as Knuth's linking automata [23], in 1977 as Tarjan's Reference Machines [35], in 1970th as Schönhage's storage modification machines [30, 31]. The evolution — with its main outcome in relaxing condition of bounded in- and out-degrees of the storage graph — ultimately reached the state of Random Access Machines, architectures of modern computers.

KUM are defined on a colored/labeled undirected graph with bounded degrees of nodes and bounded number of colors/labels. KUM operates, modify their storage, as follows. Select an active node in the storage graph. Specify local active zone, the node's neighborhood. Modify the active zone: add new node with the pair of edges, connecting the new node with the active node; delete a node with the pair of incident edges; add/delete edge between the nodes. A program for KUM specifies how to replace neighborhood of active node with new neighborhood, depending on labels of edges connected to the active node and labels of the nodes in proximity of the active node [11].

What is about computational power of KUM? Turing and KU machines compute the same class of functions. Functions computable on Turing machines are computed in on KUM and *vice verse*. Any sequential device can be simulated by KUM. Essentially any computation performing only one local action at a time is simulated by KUM [20]. However, KUM is more flexible then Turing machine because KUM is not tightened to a fixed topology of the storage space. KUM can be seen as TM which changes topology of its tape during computation.

In 1976 Grigoriev [19] demonstrated that there are predicates recognizable in real time by KUM but not recognizable in real time by Turing machine. As Cloteaux and Ranjan state [15], KUM is stronger than any model of com-putation that requires $\Omega(n)$ time access its memory. They also point out to results by Shvachko [34], who proved that KUM is stronger then any 'tree' machine (Turing machine with tapes which are infinite-length binary trees). A multi-tape Turing machine with space complexity $P$ can be simulated by Storage Modification Machine (SMM), successor of KUM, with $O(P/logP)$ nodes in real time [12]. SMM is unlikely to be simulated in real time by KUM [20], because SMM has unbounded degrees of nodes.

Are there any real-life implementations of KUM? None has been done so far. In 1988 Gurevich [20] suggested that edge of the 'Kolmogorov complex' is not informational but also physical entity and reflects physical proximity of the nodes (thus e.g. even in three-dimensional space number of neighbors of each node is polynomially bounded). What would be the best natural implementation of KUM? Reaction-diffusion chemical computers [1, 3] lack flexibility, and stationary or dissipative structures formed by them, are rather static or quasi-static. DNA and other molecular computers have almost no structure, and acts mainly as bulk, well-stirred, media computing devices. A potential candidate should be capable for growing, unfolding, graph-like storage structure, dynamically manipulating nodes and edges, and should have a wide range of functioning parameters. Vegetative stage, plasmodium, of a true slime mold *Physarum polycephalum* satisfies all these requirements.

## 2 Plasmodium of true slime mold

Plasmodium, a vegetative stage of true slime mold *Physarum polycephalum*, is a huge, and visible by naked eye, single cell (with thousands of diploid nuclei) formed when individual flagellated cells or amoebas of *Physarum polycephalum* swarm together and fuse. When placed on an appropriate substrate plasmodium propagates searching for sources of nutrients. When such sources are located and taken over, the plasmodium forms characteristic veins of protoplasm, which contracts periodically. The veins can branch, and eventually the plasmodium spans the space with a endlessly changing proximity graph, resembling but not perfectly matching graphs from the family of $k$-skeletons [22]. Essentially, plasmodium behaves as a reaction-diffusion system, enclosed in a membrane [5].

Plasmodium's ability to perform useful computational tasks, in its propagating and foraging behavior, was firstly publicized by Toshiyuki Nakagaki and his colleagues [28, 27, 29]. The plasmodium of true slime mold is proved to be a unique, user friendly and phenomenologically rich object, capable to support varieties of non-classical computation schemes [7, 8, 39], including shortest path [28, 27, 29] and robot navigations [40]. In the next section we will show how to exploit and utilize unique features of the plasmodium in an attempt to realize KUM.

# 3 Physarum machines

In the section we provide step-by-step comparison of KUM and plasmodium of true slime model, and demonstrate on experimental examples, that plasmodium of *Physarum polycephalum* is an almost ideal biological implementation of KUM.

**Methods and techniques:** The scoping experiments were designed as follows. We either covered container's bottom with a piece of wet filter paper and placed a piece of living plasmodium[1] on it, or just planted plasmodium on a bottom of a bare container and fixed wet paper on the container's cover to keep humidity high. Oat flakes were distributed in the container to supply nutrients and represent part, or data-nodes, of Physarum machine. The containers were stored in the dark except during periods of observation. To color oat flakes, where required, we used SuperCook Food Colorings[2]: blue (colors E133, E122), yellow (E102, E110, E124), red (E110, E122), and green (E102, E142). Flakes were saturated with the colorings, then dried.

**Nodes:** Physarum machine has two types of nodes: stationary nodes, presented by sources of nutrients (oat flakes), and dynamic nodes, sites where two or more protoplasmic veins originate (Fig. 1). At the beginning of computation, stationary nodes are distributed in the computational space, see example in Fig. 2, and plasmodium is placed at one point of the space. Starting in the initial conditions the plasmodium exhibits foraging behavior, and occupies stationary nodes.

**Edges:** An edge of Physarum machine is a strand, or vein, of protoplasm connecting stationary and/or dynamic nodes. KUM machine is an undirected graph, i.e. if nodes $x$ and $y$ are connected then they are connected by two edges $(xy)$ and $(yx)$. In Physarum machine this is implemented by a single edge but with periodically reversing flow of protoplasm [21, 26]. The protoplasm stream runs with speed circa 1-3 mm/s [38], or sometimes up to 4 mm/s [21], changing direction every 1–3 min. This may be also associated with periodic inversion of the potential.

**Data, results and halting:** Program and data are represented by a spatial configuration of stationary nodes. Result of the computation over stationary data-node is presented by configuration of dynamics nodes and edges. The initial state of a Physarum machines, includes part of input string (the part which represents position of plasmodium relatively to stationary nodes), empty output string, current instruction in the program, and storage structure consists of one isolated node.

What is a result of computation? Kolmogorov and Uspensky wrote: "If $S$ is a terminal state, then the connected component of the initial vertex is considered to be the "solution"" [25]. That is the whole graph structure developed by plasmodium is the result of its computation.

---

[1] Thanks to Prof. Soichiro Tsuda for providing me with *P. polycephalum* culture.
[2] `www.supercook.co.uk`

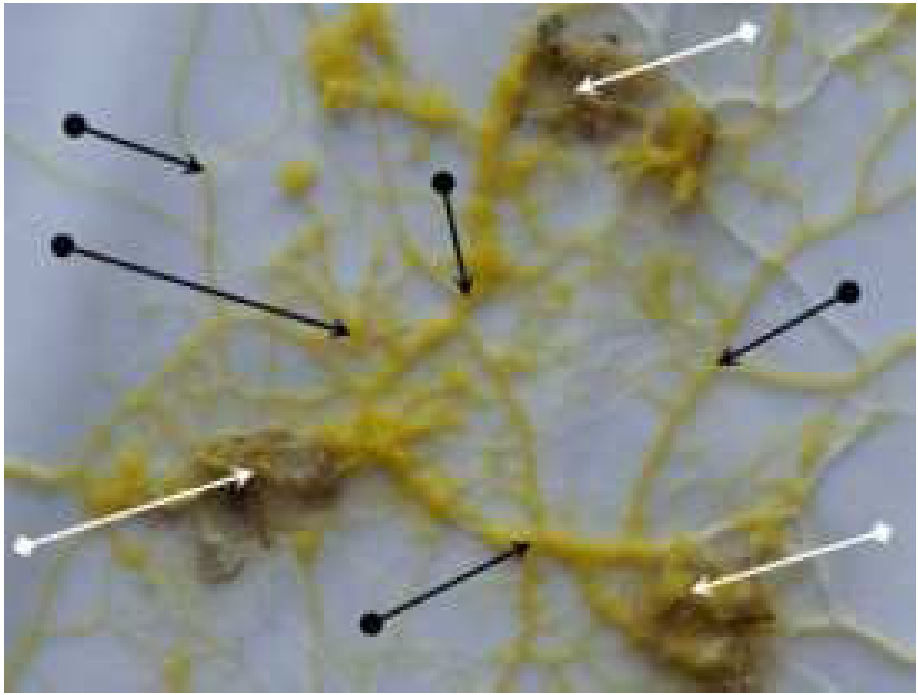

Figure 1: Nodes of Physarum machine. Stationary nodes are indicated by white arrows with rhomboid beginnings, dynamic nodes by black arrows with discoidal beginnings.

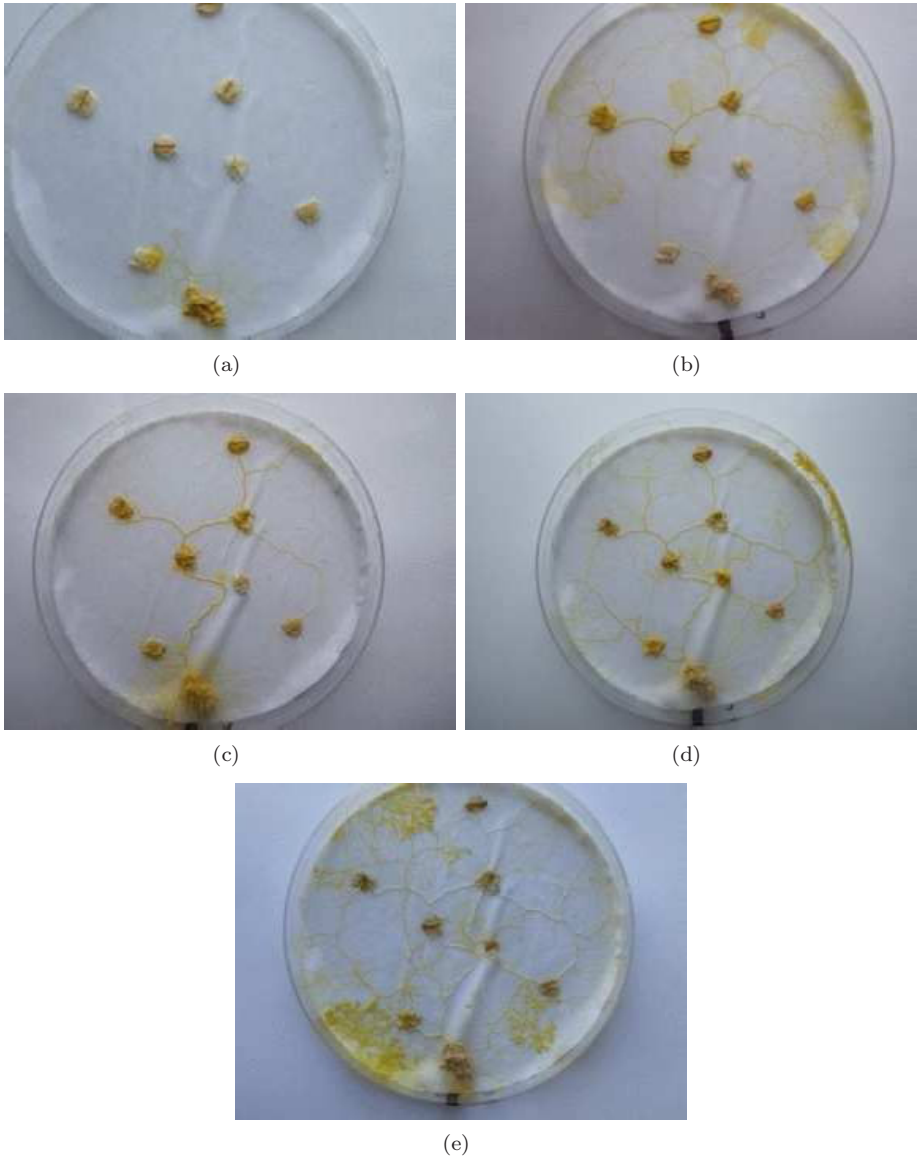

Figure 2: An example of computational process in Physarum machine. Photographs (a)–(h) are taken with time lapse circa 24 hours.

When discussing KUM, Blaas and Gurevich [11] point that "the process runs until either the next step is impossible or a signal says the solution has been reached". Plasmodium proceeds computation even when solution has been reached and halts only when physical resources are exhausted. Plasmodium continues its spreading, reconfiguration and development till there are enough nutrients. When supply of nutrients is over, plasmodium either switches to fructification state (if level of illumination is high enough), when sporangia are produced, or forms sclerotium (encapsulates itself in hard membrane), if in darkness. Therefore, we propose that Physarum machine halts when all data-nodes are utilized.

Formation of sclerotium can be seen as freezing of the storage structures, no nodes added or deleted, edges remain stationary. Fructification type halting is equivalent to self-reproduction of Physarum machine.

**Active zone:** In KUM storage graph must have some active node. This is an inbuilt feature of Physarum machine. When plasmodium resides on a substrate with poor or no nutrients, then just one or few nodes generate actively spreading protoplasmic waves. In these cases the protoplasm spreads as mobile localizations, or localized wave-fragments, by analogy with wave-fragments in sub-excitable Belousov-Zhabotinsky medium [32]. An example of single active node, which is just started to develop its active zone, is shown in (Fig. 3).

At every step of computation in KUM there is an active node and some active zone, usually nodes neighboring to active node. The active zone has limited complexity, in a sense that all elements of the zone are connected by some chain of edges to the initial node.

In general, size of active zone may vary depending on computational task. In Physarum machine an active node is a trigger of contraction/excitation waves, which spread all over the plasmodium tree and cause pseudopodia to propagate, shape to change and even protoplasmic veins to annihilate. Active zone is comprised of stationary or dynamic nodes connected to active node with veins of protoplasm.

**Bounded connectivity:** In contrast to Schönhage machine KUM has bounded in- and out-degree of the storage graph. Graphs developed by Physarum are predominantly planar graphs. Moreover, if we put a piece of vein of protoplasm on top of another vein of protoplasm, the veins fuse [33]. Usually, not more then three protoplasmic strands join each other in one given point of space. It is reported that average degree of minimum spanning tree is around 1.99, and of relative neighborhood graph around 2.6 [14]. Graphs produced by standard procedures for generating combinatorial random planar graphs show a limited growth of average degree with number of nodes or edges, the degree stays around 4 when number of edges increase from 100 to 4000 [6]. We could assume that average degree of storage graph in Physarum machines is a bit higher then degree of spanning trees but less then degree of random planar graphs.

**Addressing and labeling:** Every node of KUM must be uniquely addressable and nodes and edges labeled, "the vertices joined by edges to a fixed vertex belong to different types" [25]. There is no direct implementation of such addressing in Physarum machine. With stationary nodes this can be implemented either by coloring the nodes, or by tuning humidity of the oat flakes, thus changing concentration of bacteria living on the flakes. Coloring the stationary nodes could be another solution. Few examples of experimental implementation of addressing and labeling are shown in Fig. 4. Addressing green colored flake is shown in Fig. 4a, and addressing blue flake with green-colored edge in Fig. 4b. Plasmodium does not treat all colors the same. So far we observed the following priority of choosing colored flakes, in descending order: uncolored flakes, green colored, yellow and blue colored, and red colored. The preferences allow plasmodium to selectively connect labeled nodes in the storage graph, an example is demonstrated in Fig. 4cd.

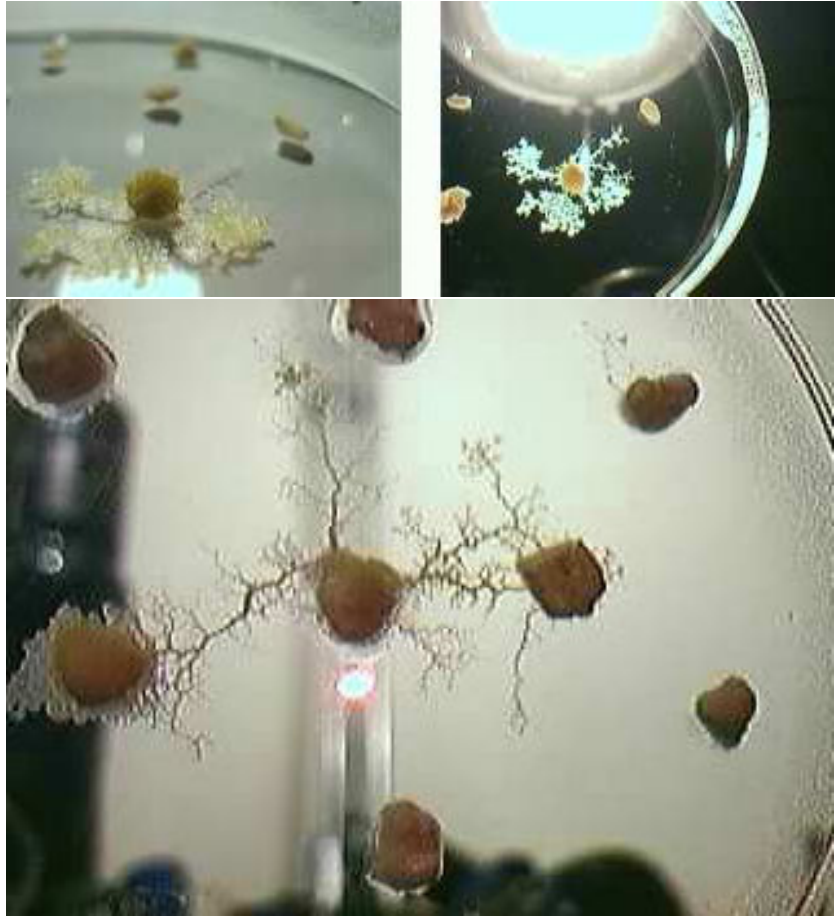

Figure 3: Active node and active zone: (a) single active node is generating active zone at the beginning of computation, (b) active zone around a central active node consists of two nodes.

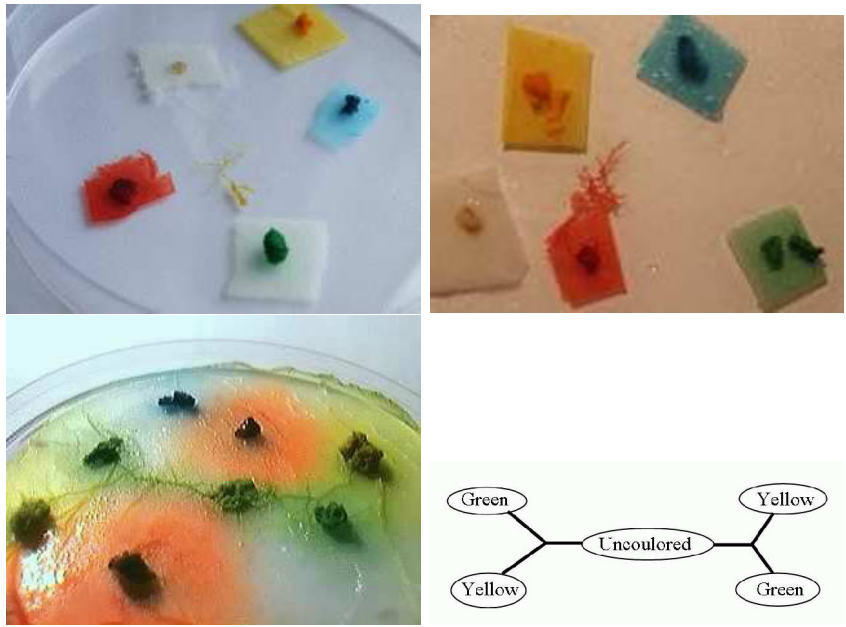

Figure 4: Addressing and labeling in Physarum machine: (a) machine is addressing green-colored data-node, (b) edge outcoming one data-node is labeled in red, (c) plasmodium selectively connects green and yellow nodes in a storage structure, (d) schema of the connection (c).

**Basic operations:** A possible set of instructions for Physarum machine could be as follows. Common instruction would include INPUT, OUTPUT, GO, HALT, and internal instructions: NEW, SET, IF [16]. At present state of experimental implementation we assume that INPUT is done via distribution of sources of nutrients, while OUTPUT is recorded optically. Halting was discussed in previous sections. SET instruction causes pointers redirection, and can be realized by placing fresh source of nutrients in the experimental container, preferably on top of one of the old sources of nutrients. When new node is created all pointers from the old node point to the new node.

Let us look at the experimental implementation of core instructions.

ADD NODE: To add a stationary node $b$ to node $a$'s neighborhood, plasmodium must propagate from $a$ to $b$, and form a protoplasmic vein representing edge $(ab)$. To form a dynamic node, propagating pseudopodia must branch into two or more pseudopodia, and the site of branching will represent newly formed node, see example in Fig. 1.

REMOVE NODE: To remove stationary node from Physarum machine, plasmodium leaves the node (Fig. 5). Annihilating protoplasmic strands forming a dynamic node at their intersection, remove the dynamic node.

ADD EDGE: To add an edge to a neighborhood, active node generates propagating processes which establish a protoplasm vein with one or more neighboring nodes. See formation of protoplasmic veins between nodes marked by solid discs in Fig. 5.

REMOVE EDGE: When protoplasmic vein annihilates, e.g. depending on global state or when source of nutrients exhausted, edge represented by the vein is removed from Physarum machine (Fig. 6). The following sequence of operations is demonstrated in Fig. 6: node 3 is added to the structure by removing edge (12) and forming two new edges (13) and (23).

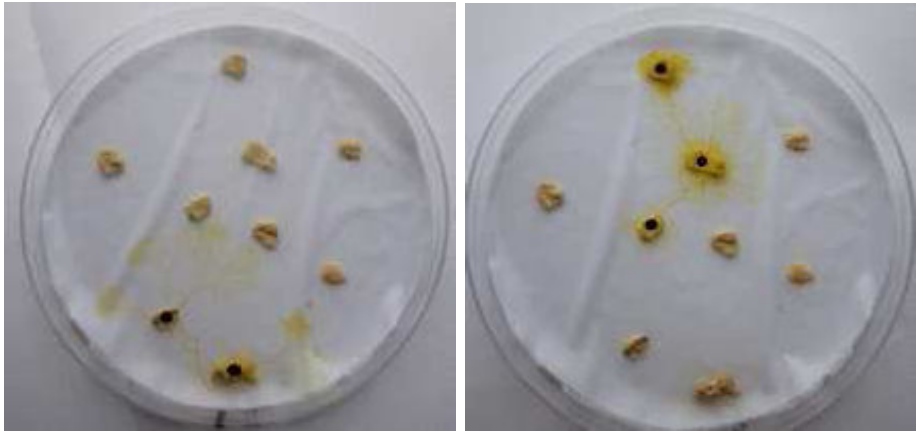

Figure 5: Implementation of REMOVE NODE by Physarum machine. At first plasmodium occupies two stationary nodes (a) at the south of the Petri dish, then removes these nodes from its storage structure by leaving them and occupying three stationary nodes at the north part of the Petri dish (b). Nodes included in the storage structure are marked by solid discs.

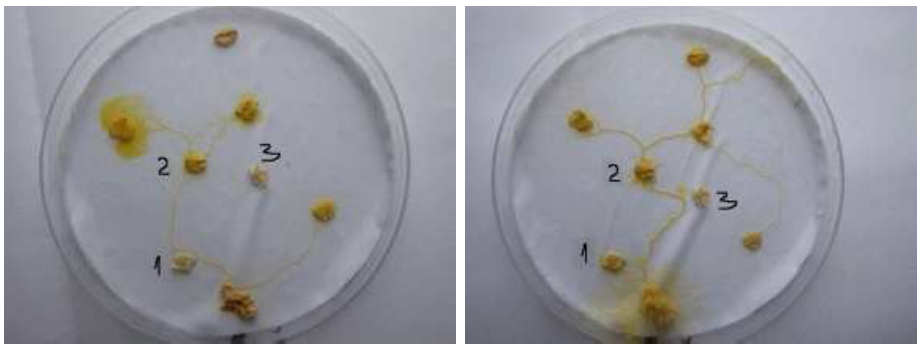

Figure 6: Reconfiguration of storage structure by Physarum machine, implementation of ADD NODE, ADD EDGE, REMOVE EDGE operations.

# 4 Discussions

We have suggested a possible biological substrate for realization of Kolmogorov-Uspensky machine (KUM). In the experimental implementation we exploited the fact that in contrast to Turing machine (TM), which reads and react on symbols written on the tape, KUM acts depending on the state of neighborhood and its currently active node, includes neighboring nodes and edges. Despite overwhelming popularity of Turing machines, modern computing devices evolved from KUM to Schönhage Storage Modification Machines and other pointer machines, and then to Random Access Machines.[3] We can speculate therefore that Physarum machines are closest ever biological realizations of nowadays general-purposed computers.

There are several issues which remain unresolved, or not addressed in full details in present paper.

Thus, we did not show exactly how to satisfy Kolmogorov property of the storage graph – each node is uniquely addressable. However we provided possible experimental implementations of the addressing via coloring of stationary nodes.

There is also an important question of parallelism. KUM, in its original form, has a single control device, at least 'modern age' interpretations [20, 11] claim so. Does it apply to Physarum machines? Partly yes. Plasmodium acts as a unit in a long-term, i.e. it can change position, or retract some processes in one place to form new ones in another place. However, periodic contractions of protoplasm are usually initiated from a single source of contraction waves [26, 36]. This source of waves can be interpreted as a single control unit. Said that, during foraging behavior, several branches or processes of plasmodium can act independently and almost autonomously.

When thinking about parallelism in Physarum machine, we can recall Bardzin's Growing Automata, outlined in 1964 [9], and developed by Bardzin and Kalnins in details [10] (these are also utilized in casual networks [18]). Bardzin Growing Automata (BGA) are — in reality — parallel extensions of KUM. In these automata local transformations are simultaneously implemented in all nodes (and primitive operations are similar to that in KUM).

A possible compromise between original theoretical framework of KUM and partly parallel execution in experiments could be reached by proposing two levels of 'biological' commands executed by Physarum machine's elements. They are high-level commands, e.g. SEARCH FOR NUTRIENTS, ESCAPE LIGHT, FORM SCLEROTIUM, FRUCTIFY, and low-leve commands, e.g. FORM PROCESS, PROPAGATE IN DIRECTION OF, OCCUPY SOURCE OF NUTRIENTS, RETRACT PROCESS, BRANCH. Global commands are executed by plasmodium as a whole at once, i.e. in a given time step plasmodium executes only one high-level command. Local commands are executed by local parts of the plasmodium. Two spatially

---

[3]Possibly this happens because Kolmogorov and Uspensky clearly shifted focus from *algorithm* to *machine*: "...the schema for computing the value of a partially recursive function may not be directly given in the form of algorithm. If one develops this computation in the form of an algorithmic process.... then, through this, one automatically obtains a certain algorithm..." [25].

distant sites can execute different low-level commands.

# Acknowledgment

My sincere thanks go to Drs. Soichiro Tsuda and Tomohiro Shirakawa who ignited my interest in experimenting with plasmodium and were my patient advisers.